\documentclass[review]{elsarticle}

\usepackage[fleqn]{amsmath}
\usepackage{lineno,hyperref}
\modulolinenumbers[5]

\journal{Journal of \LaTeX\ Templates}

\makeatletter
\def\@author#1{\g@addto@macro\elsauthors{\normalsize%
    \def\baselinestretch{1}%
    \upshape\authorsep#1\unskip\textsuperscript{%
      \ifx\@fnmark\@empty\else\unskip\sep\@fnmark\let\sep=,\fi
      \ifx\@corref\@empty\else\unskip\sep\@corref\let\sep=,\fi
      }%
    \def\authorsep{\unskip,\space}%
    \global\let\@fnmark\@empty
    \global\let\@corref\@empty  
    \global\let\sep\@empty}%
    \@eadauthor={#1}
}
\makeatother
\begin{document}

\begin{frontmatter}

\title{Kinetic theory of binary particles with unequal mean velocities and  non-equipartition energies}

\author{Yanpei Chen\corref{mycorrespondingauthor}}
\cortext[mycorrespondingauthor]{Corresponding authors}
\ead{ypchen@ipe.ac.cn}
\author{Yifeng Mei}
\author{Wei Wang\corref{mycorrespondingauthor}}
\ead{wangwei@ipe.ac.cn}


\address{State Key Laboratory of Multiphase Complex Systems, Institute of Process Engineering, Chinese Academy of Sciences, Beijing 100190, China}
%
%

\begin{abstract}
The hydrodynamic conservation equations and constitutive relations for a binary granular mixture composed of smooth, nearly elastic spheres with non-equipartition energies and different mean velocities are derived. This research is aimed to build three-dimensional kinetic theory to characterize the behaviors of two species of particles suffering different forces. The standard Enskog method is employed assuming a Maxwell velocity distribution for each species of particles. The collision components of the stress tensor and the other parameters are calculated from the zeroth- and first-order approximation. Our results demonstrate that three factors, namely the differences between two granular masses, temperatures and mean velocities all play important roles in the stress-strain relation of the binary mixture, indicating that the assumption of energy equipartition and the same mean velocity may not be acceptable. The collision frequency and the solid viscosity increase monotonously with each granular temperature. The zeroth-order approximation to the energy dissipation varies greatly with the mean velocities of both species of spheres, reaching its peak value at the maximum of their relative velocity.
\end{abstract}

\begin{keyword}
kinetic theory of granular flow; binary granular mixture; energy non-equipartition
\MSC[2010] 00-01\sep  99-00
\end{keyword}

\end{frontmatter}

\section{\label{sec:level1}Introduction}
Granular mixtures \cite{ISI:A1990CK57100004,goldhirsch2003rapid,ISI:000220892300001}, including e.g., landslides, avalanches and pharmaceutical powders, are common in nature and industry. Granular mixtures and ordinary molecular gas mixtures act differently \cite{Torquato2010Jammed,Donev2005Pair,ISI:000237893800013,ogarko2013prediction}. For a granular mixture particularly engaged in rapid flow, nearly instantaneous and inelastic collisions happen between particles, leading to dissipation. So it is necessary to inject energy continuously to keep a steady state for a granular mixture. Two ways of energy compensation are usually involved:  boundary driving, as vibration \cite{ISI:000220892300001,ISI:A1987G573900020} and shear \cite{ISI:A1981MP06200015}, and bulk driving, as air-fluidization \cite{Gidaspow1,ISI:A1990CY99300003} and magnetic field \cite{PhysRevLett.94.108002}. Similar to a single species granular gas \cite{PhysRevE.88.052204,yan2012breakdown}, granular mixtures are normally in non-equilibrium states. And it has been demonstrated that the components in a granular mixture do not share the same granular temperature \cite{ISI:000175324900051,ISI:000229749500015,ISI:000178537000033,ISI:000173849300019,PhysRevLett.100.158001}.

 The dense gas kinetic theory of Chapman \cite{Chapman} has been employed to quantify rapid flow of granular mixture systems for decades \cite{PhysRevE.65.061302,PhysRevLett.88.044301,PhysRevE.69.041302,ISI:000237893800013}, in which granular mixtures are assumed to be smooth, nearly elastic and spherical grains. In literature, two approaches can be classified, as elaborated by Galvin et al. \cite{ ISI:000228770800009}: the first one is derived via systematic expansion, such as Chapman-Enskog method, and the second one is on the base of a hypothesized velocity distribution. According to the factors of the velocity distribution (Maxwellian or non-Maxwellian velocity distribution), equipartition or non-equipartition (equal or unequal granular temperature), radial distribution(standard Enskog or revised Enskog theory \cite{ISI:A1987G582600005}), system dimension (two or three dimensions), the second approach can be further divided. Jenkins and Mancini \cite{ISI:A1987G582600005} firstly derived constitutive relations and balance equations by employing the Enskog equations with the assumption that the velocities of two species of particles are Maxwellian distributed, and the temperatures of two species were supposed to differ by infinitely small quantity. Based on the revised Enskog theory \cite{ISI:A1989CC47300017,ISI:A1995TG33600041,ISI:000075379000008}, more exact theories with equipartition assumption were further developed, concerning the non-uniform, local equilibrium sate. The kinetic theory of non-equipartition, binary granular mixtures with different sizes, masses and diameters was also obtained based on the Enskog equations by Lu et al. \cite{PhysRevE.64.061301,ISI:000187361700002}, where two granular temperatures were assumed, respectively, for different sized particles. However, both Rahaman et al. \cite{ISI:000187361700002} and  Iddir \& Arastoopour \cite{ISI:000229306300005} pointed out that Lu's collision rate between particles $i$ and $j$, $N_{ij}$, is not symmetric, that is, $N_{ij}\neq N_{ji}$. Rahaman et al. \cite{ISI:000187361700002} improved the integration processes by supposing the angle between the relative and combined velocities of two species of particles $i$ and $j$ is in the range of $[0, 2\pi]$, implying both of these velocities are two-dimensional vectors. That treatment contradicts the three-dimensional derivation in their work. Moreover, all the above derivations supposed that the mean velocities of two species are identical, which is obviously not the case in gas-fluidized systems \cite{JinghaiLi,ISI:000243095400018,ISI:000321964700024,ISI:000344943800010}. The first approach developed by Garz\'{o} et al. \cite{ISI:000174319200021,ISI:000249785800054,ISI:000366207400011} was dedicated to solving perturbatively the Boltzmann equations to capture a broader range of restitution. However, Galvin et al. \cite{ISI:000228770800009} pointed out that the application of the second approach could be extended to moderately dense system, whereas the first one is limited to dilute granular gases. Recently Garz\'{o} et al. \cite{ISI:000249785800055,ISI:000302338900005} extended the first approach to the (moderately) dense flow. Serero et al.\cite{FLM:9976180} further provided a hydrodynamic description of dilute binary gas mixtures comprising smooth inelastic spheres interacting by binary collisions with a random coefficient of restitution. Both of these approaches have been used to model the stress of particles in multiphase computational fluid dynamics (CFD) \cite{Gidaspow1,ISI:000299404900034, ISI:000253798000017,ISI:000184851900018, ISI:000278060600009}.

  It is worth noting that not only the binary granular mixtures but the air-fluidized granular systems with identical spheres are also in non-equilibrium states \cite{ISI:A1987G582600005,PhysRevE.64.061301, ISI:000187361700002,ISI:000249785800054}. In gas-solid circulating fluidized beds, dense clusters \cite{ISI:A1979GH83500007,ISI:000171670500007} with size of $10$-$100$ times the particle diameter are suspended in the dilute broth of gas-solid mixture, whereas the particles accumulated in the cluster and those dispersed in the dilute broth do not share the granular temperature and mean solid velocity \cite{ISI:000171670500007}. This is reasonable because the particles in the dense cluster and dilute both suffer different drag forces \cite{JinghaiLi,ISI:000344943800010} which make the particles away from equilibrium states (share the same granular temperature). To illustrate this inhomogeneous structure, it is convenient to elucidate the above situation by using two velocity distributions with two temperature equations. Francisco et al. \cite{reyes2007granular}  presented a granular mixture model for elastic spheres subject to drag force with different mean velocities which cannot describe dissipation. Thus, the kinetic theory \cite{3035943} of double granular temperatures and mean velocities makes sense not only for the binary granular mixtures but also for CFD simulation of fluidization\cite{ISI:000253798000017,ISI:000278060600009,ISI:000089464700006,Lu2007investigation}.

In this paper we dedicate to deriving the constitutive relations and balance equations of two kinds of particles with unequal mean velocities and  non-equipartition energies using the standard Enskog theory. Sec. \ref{sec:level21} provides a binary collision frequency derived through their dependence on the velocity distribution function of a mixture of inelastic spheres. In Sec. \ref{sec:level25}, the hydrodynamic descriptions are given based on the Boltzmann equation. In Sec. \ref{sec:level31}, the constitutive equations are identified by macroscopic hydrodynamic variables. In Sec. \ref{sec:level41} we discuss our results and compare with previous works. In Sec. \ref{sec:level61}, the main conclusions are summarized.

%
%
%
\section{\label{sec:level21}Binary collision frequency}
For a binary granular mixture composed of smooth, inelastic spheres of species $\alpha$ and $\beta$ with mass $m_{i}$ and diameter $d_{i}$, $i=\alpha$,$ \beta$, due to suffering inhomogeneous external energy input, such as the different drag forces exerted on fluidized particles, two species may have their own temperatures and mean velocities, respectively. Here, we consider the most general case for binary collisions.
Let $i$ and $j$ represent either species $\alpha$ or $\beta$.
For an inelastic collision between two particles in species $i$ and $j$, with velocity $\textbf{c}_i$ and $\textbf{c}_j$, the relationship between pre- and post- collision relative velocities $\textbf{c}_{ij}=\textbf{c}_{i}-\textbf{c}_{j}$ and $\textbf{c}_{ij}'=\textbf{c}_{i}'-\textbf{c}_{j}'$ yields
\begin{equation}
\label{collisionrelation}
\begin{aligned}
&\textbf{k} \cdot \textbf{c}_{ij}'=-e_{ij}(\textbf{k}\cdot \textbf{c}_{ij})
\end{aligned}
\end{equation}
where $\textbf{k}\equiv \textbf{r}_{ij}/r_{ij}$ is the unit vector directing from the center of particle with velocity $\textbf{c}_i$ to that of particle with velocity $\textbf{c}_j$ upon contact, specifying the geometry of the impact. $e_{ij}$ is the restitution coefficient between species $i$ and $j$. Angular velocities or rotations are not included in this paper. The velocity of the center of mass $\emph{\textbf{G}}$ is defined as:
\begin{equation}
\emph{\textbf{G}}=\frac{m_i \emph{\textbf{c}}_i +m_j \emph{\textbf{c}}_j}{m_0}
\end{equation}
where $m_{0}=m_{i}+m_{j}$.  At the moment of collision, the distance between two centers of particles is $d_{ij}=(d_i+d_j)/2$.
The number of binary collisions between species $i$ and $j$ at position $\textbf{r}$ per unit time per unit volume has the form:
\begin{equation}
N_{ij}=
\int_{\textbf{c}_{ij} \cdot \textbf{k} >0}
f_{ij}^{(2)} \left(  \textbf{c}_i,  \textbf{r},  \textbf{c}_j,  \textbf{r}+d_{ij} \textbf{k}\right)
\left( \textbf{c}_{ij}  \cdot \textbf{k} \right) d_{ij}^2
\rm{d} \textbf{k} \rm{d} \textbf{c}_i \rm{d} \textbf{c}_j
\end{equation}
where $f_{ij}^{(2)}$ is the pair distribution function. Following Chapman and Cowling \cite{Chapman}, the assumption of chaos allows us to write the correlation probability function as a product of two single velocity distributions:
\begin{equation}
f_{ij}^{(2)}\left(\textbf{c}_i, \textbf{r} , \textbf{c}_j , \textbf{r}+d_{ij} \textbf{k} \right)=\chi (\textbf{r}+\frac{1}{2}d_{ij} \textbf{k})f_i\left(\textbf{c}_i,\textbf{r}\right)f_j\left(\textbf{c}_j,\textbf{r}+d_{ij} \textbf{k}\right)
\end{equation}
where the factor $\chi$ called Enskog factor equals to unity for a rare gas \cite{Chapman}, and reads \cite{ISI:000229306300005}:
\begin{equation}
 \chi (\textbf{r}+\frac{1}{2}d_{ij} \textbf{k})=g_{ij}=
 \frac{d_jg_{ii}(\varepsilon_i,\varepsilon_j)+d_ig_{jj}(\varepsilon_i,\varepsilon_j)}{2d_{ij}}
\end{equation}
in which
\begin{equation}
 g_{ii}(\varepsilon_i,\varepsilon_j)=
 \frac{1}{1-(\varepsilon_i+\varepsilon_j)/\varepsilon_{max}}+
 \frac{3d_i}{2}
 \sum_{k=i,j}\frac{\varepsilon_k}{d_k}
\label{genskog}
\end{equation}
And $\varepsilon_i$ and $\varepsilon_j$ are solid volume fraction for species $i$ and $j$, respectively, $\varepsilon_{max}$ is the single-phase maximum packing, $g_{ij}$ is the radial distribution function between spheres of species $i$ and $j$.

And we assume the velocities of both species of particles follow the Maxwellian distribution:
\begin{equation}
f_{i}\left(\textbf{c}_{i},\textbf{r}\right)=n_i \left( \frac{m_{i}}{2\pi \theta_{i}}\right)^{3/2}
\exp \left[-\frac{m_{i}\left(\textbf{c}_{i}-\textbf{v}_{i}\right)^2}{2 \theta_{i}}\right]
\end{equation}
where $n_i$ is the particle number density, $\textbf{v}_{i}$ is the mean velocity of species $i$, and $\theta_i$ is the granular temperature defined as an ensemble average $\theta_{i}=\frac{1}{3}m_i<(\textbf{c}_{i}-\textbf{v}_{i})^2>$.

For the binary mixture, the mean mass center velocity is defined by
\begin{equation}
\textbf{v}=\frac{n_{i}m_{i} \textbf{v}_{i}+n_{j}m_{j} \textbf{v}_{j}}{n_{i}m_{i}+n_{j}m_{j}}
\end{equation}
The peculiar velocity is the relative velocity between the particle velocity and the mean velocity£º
\begin{equation}
\textbf{C}_{i}\equiv \textbf{c}_{i}-\textbf{v}_i
\end{equation}
The diffusion velocity $\textbf{u}_{i}$ reads
\begin{equation}
  \textbf{u}_{i} \equiv \textbf{v}_{i}-\textbf{v}
\end{equation}
which denotes the mean velocity of species $i$ relative to the mean mass center velocity.

Then the joint pair distribution function becomes
\begin{equation}
\begin{aligned}
f_{ij}^{(2)}\left(\textbf{c}_i,\textbf{r}_i,\textbf{c}_j,\textbf{r}_j\right)
=&\frac{1}{8 \pi^3} g_{ij} n_i n_j\left(\frac{m_i m_j}{\theta _i \theta _j}\right)^{3/2}
 \exp\left[-\frac{m_i\left(\textbf{c}_i-\textbf{v}_i\right)^2}{2 \theta _{i}}-\frac{m_{j}\left(\textbf{c}_j-\textbf{v}_j\right)^2}{2 \theta _j}\right]
\label{pairdistirbution}
\end{aligned}
\end{equation}
Using this pair distribution, the collision frequency becomes
\begin{equation}
\begin{aligned}
N_{ij}=\frac{1}{8 \pi^2}&d_{ij}^2 g_{ij} n_i n_j
\left( \frac{m_i m_j}{\theta_i \theta_j}\right)^{3/2}\\ &\iint c_{ij}
\exp\left[-\frac{m_i\left(\textbf{c}_i-\textbf{v}_i\right)^2}{2 \theta_i}-\frac{m_j\left(\textbf{c}_j-\textbf{v}_j\right)^2}{2 \theta _j}\right]
\rm{d}\textbf{c}_i \rm{d}\textbf{c}_j
\end{aligned}
\end{equation}

We can express the particles velocities in terms of \emph{\textbf{G}}, $\textbf{c}_{ji}$ as in literature \cite{PhysRevE.64.061301,ISI:000187361700002}, by expanding it in a Taylor series:
\begin{equation}
\begin{aligned}
\label{frequencyExp}
N_{ij}=&\frac{1}{8 \pi^2}d_{ij}^{2}g_{ij}n_{i}n_{j}\left(\frac{m_i m_j}{\theta _i \theta _j}\right)^{3/2}\iint c_{ij}\exp[-A\textbf{G}^2-D\textbf{c}_{ji}^{2}] \\
&\times [1-2B(\textbf{G}\cdot \textbf{c}_{ji})+2B^{2}(\textbf{G}\cdot \textbf{c}_{ji})^{2}+...]\rm{d} \textbf{G}\rm{d} \textbf{c}_{ji}
\end{aligned}
\end{equation}
where

$$
A=\frac{m_i\theta_j + m_j\theta_i}{2\theta_i\theta_j}, B=\frac{ m_i m_j(\theta_i-\theta_j) }{2m_0\theta_i\theta_j},
$$

$$
D=\frac{ m_i m_j(m_i\theta_i+m_j\theta_j) }{2m_0^2\theta_i\theta_j}
$$

\begin{figure}
\centering
 \includegraphics[width=0.6\textwidth]{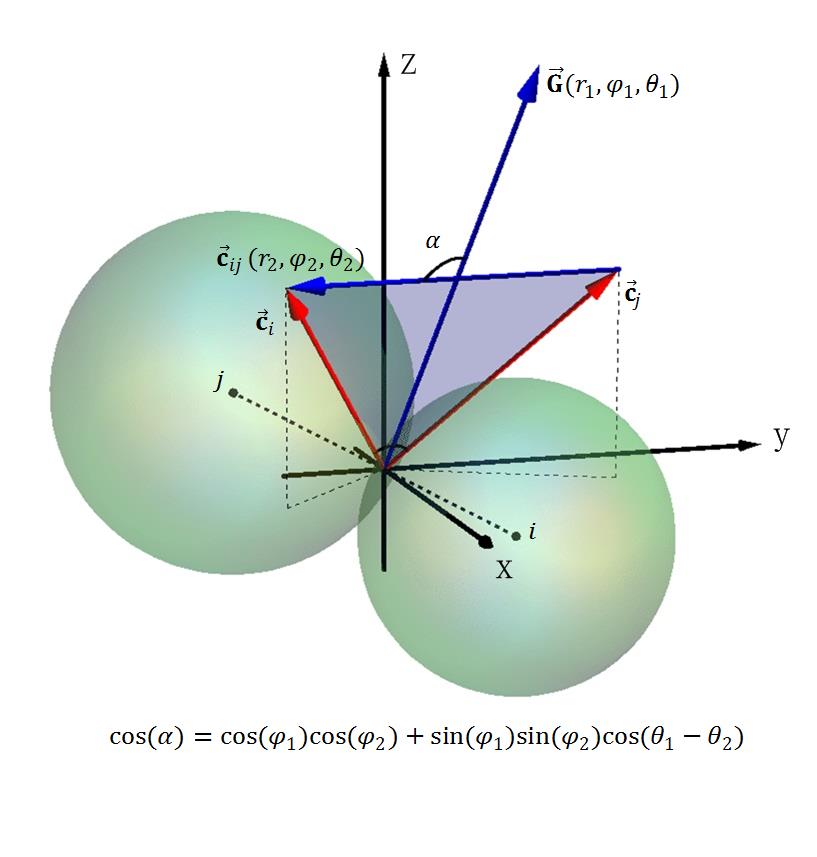}
 \caption{The sketch of a binary collision.}
 \label{sketch}
 \end{figure}
The integration of the cross term $\textbf{G}\cdot\textbf{c}_{ij}$ needs to be stressed here. In Lu's model \cite{PhysRevE.64.061301}, the mass center velocity $\textbf{G}$ and relative velocity $\textbf{c}_{ij} $ in the cross term are treated to be scalar. In Rahaman's model \cite{ISI:000187361700002}, $\textbf{G}$ and $\textbf{c}_{ij} $ are assumed in a two-dimensional plane,  so the angle between $\textbf{G}$ and $\textbf{c}_{ij} $ are in the range of $[0,2\pi]$. However, as illustrated in Fig. (\ref{sketch}), in three-dimensional spherical coordinates, the angle between two vectors $\textbf{G}(r_{1},\varphi_{1},\theta_{1})$, $\textbf{c}_{ij}(r_{2},\varphi_{2},\theta_{2})$ is
\begin{equation}
\label{3Dinsideangle}
\alpha=\arccos(\cos \varphi_{1} \cos \varphi_{2} +\sin \varphi_{1} \sin \varphi_{2} \cos(\theta_{1}-\theta_{2}))
 \end{equation}
Integrate Eq. (\ref{frequencyExp}) using Eq. (\ref{3Dinsideangle}), we obtain
\begin{equation}
N_{ij}=\frac{1}{4}g_{ij}d_{ij}^2 n_i n_j
\left(\frac{m_i m_j}{\theta _i \theta _j}\right)^{3/2}
\frac{ \sqrt{\pi} }{A^{3/2}D^2 }
\left[ 1+\frac{2B^2}{A D}+\frac{3B^4}{A^2 D^2}+\frac{4B^6}{A^3 D^3}+\ldots \right]
\label{collisionfreqresult}
\end{equation}

Due to the symmetry of collisions, the number of collisions of species $i$ is expected to equal the number  of collisions of species $j$, i.e., $N_{ij}=N_{ji}$, which implies that B in the expression of $N_{ij}$ should be raised to power of even numbers. Our results satisfy this criterion.

\section{\label{sec:level25}Conservation equations}
The velocity distribution function $f\left(\textbf{c},\textbf{r},t\right)$ of each granular species of a binary mixture subjected to an external force satisfies the Boltzmann (integro-differential) equation \cite{Chapman}
\begin{equation}
\frac{\partial f}{\partial t}+\textbf{c} \cdot \frac{\partial f}{\partial \textbf{r}}+\textbf{F} \cdot \frac{\partial f}{\partial \textbf{c}} = \left( \frac{\partial f}{\partial t} \right)_{coll}
\end{equation}
where $\left( \frac{\partial f}{\partial t} \right)_{coll}$ is the change rate of distribution function $f\left(\textbf{c},\textbf{r},t\right)$ due to particle collisions, $\textbf{F}$ represents the external force such as gravity, buoyancy and gas-solid drag force.

For species $i$, let $\psi_{i}$ be any function of particle velocity $\textbf{c}_i$. Multiply both sides of Boltzmann equation by $\psi_{i}d\textbf{c}_i$  and integrate them throughout the velocity-space, we could obtain the equation of change rate of the particle property, the transport equation for the quantity $\psi_{i}$:
\begin{equation}
\frac{\partial n_i <\psi_i>}{\partial t}
+\nabla \cdot \left( n_i <\textbf{c}_i\psi_i> + \sum_{j=\alpha, \beta} \textbf{P}_{c,ij} \right) =
n_i <\textbf{F}_i \cdot \frac{\partial}{\partial \textbf{c}_i} \psi_i> + \sum_{j=\alpha, \beta} \textbf{N}_{c,ij}
\end{equation}
where $\textbf{P}_{c,ij}$ is the collisional part of the stress tensor, and $\textbf{N}_{c,ij}$ the collisional source term for particles in species $i$ during the collision with particles in species $j$. Here, we use the expressions of $\textbf{P}_{c,ij}$ and  $\textbf{N}_{c,ij}$  given by Jenkins and Mancini \cite{ISI:A1987G582600005} in 1987:
\begin{equation}
\begin{aligned}
\textbf{P}_{c,i j}(\psi_{i})=-\frac{1}{2}d_{ij}^3 &\int_{\textbf{k} \cdot \textbf{c}_{ij}>0}
\left(\psi_{i}'-\psi_{i}\right)
\left(\textbf{k} \cdot \textbf{c}_{ij} \right)\textbf{k} \\
&\times f_{ij}^{(2)}\left(\textbf{c}_i,\textbf{r}-\frac{1}{2}d_{ij}\textbf{k},\textbf{c}_j,\textbf{r}+\frac{1}{2}d_{ij}\textbf{k}\right)
d\textbf{k}d\textbf{c}_id\textbf{c}_j
\label{generalPc}
\end{aligned}
\end{equation}
\begin{equation}
\textbf{N}_{c,i j}(\psi_{i})= d_{ij}^2 \int_{\textbf{k} \cdot \textbf{c}_{ij}>0}
\left(\psi_{i}'-\psi_{i}\right)
\left(\textbf{k} \cdot \textbf{c}_{ij} \right)
f_{ij}^{(2)}\left(\textbf{c}_i, \textbf{r}-\frac{1}{2}d_{ij}\textbf{k},\textbf{c}_j,\textbf{r}+\frac{1}{2}d_{ij}\textbf{k}\right)
d\textbf{k}d\textbf{c}_id\textbf{c}_j
\end{equation}
Setting $\psi_i = m_i$, we get the balance equation of mass.
\begin{equation}
\frac{\partial (n_i m_i)}{\partial t} +
\nabla \cdot (n_i m_i \textbf{v}_i)=0
\label{numberconversation}
\end{equation}
Setting $\psi_i = m_i \textbf{c}_{i}$, we get the balance equation of momentum.
\begin{equation}
\begin{aligned}
&\frac{\partial (n_i m_i \textbf{v}_i)}{\partial t} +
\nabla \cdot (n_i m_i \textbf{v}_i \textbf{v}_i) = \\
&-\nabla \cdot \left[
\sum_{j=\alpha,\beta} \textbf{P}_{c,ij}(m_i \textbf{C}_{i})+\textbf{P}_{k,i} \right]
+n_i m_i \textbf{F}_i
+ \sum_{j=\alpha, \beta} \textbf{N}_{c,ij}(m_i \textbf{C}_{i})
\label{momentconservation}
\end{aligned}
\end{equation}
Among it, $\textbf{P}_{k,i}$ is the kinetic part of the stress tensor for species $i$.
\begin{equation}
\textbf{P}_{k,i}=\int m_i \textbf{C}_i \textbf{C}_i f_i \rm{d} \textbf{c}_i
\label{PKI}
\end{equation}


Setting $\psi_i$ as $\frac{1}{2}m_i \textbf{c}_i^2$ and subtracting it by the product of the momentum balance equation (Eq. (\ref{momentconservation}))
 and hydrodynamic velocity $\textbf{v}_i$, we obtain the balance equation of granular temperature,
\begin{equation}
\begin{aligned}
\frac{3}{2} \left[
\frac{\partial}{\partial t} (n_i \theta_i) + \nabla \cdot (n_i \textbf{v}_i \theta_i) \right]=&
\left[\textbf{P}_{k,i}+\sum_{j=\alpha,\beta} \textbf{P}_{c,ij}(m_i \textbf{C}_i) \right]:\nabla \textbf{v}_i -
\nabla(\textbf{q}_{k,i}+\textbf{q}_{c,i}) \\&+
n_i m_i<\textbf{F} \cdot \textbf{C}_i >+\sum_{j=\alpha,\beta}\textbf{N}_{c,ij}
\left( \frac{1}{2}m_i \textbf{C}_i^2
\right) 
\label{energyconversation}
\end{aligned}
\end{equation}
where $\textbf{q}_{c,i}$ represents \(\sum_{j=\alpha,\beta}\textbf{P}_{c,ij}
\left( \frac{1}{2} m_i \textbf{C}_i^2  \right)\), and $\textbf{q}_{k,i}$ is the kinetic part of energy flux which is given by
\begin{equation}
\textbf{q}_{k,i}=\int \frac{1}{2}m_i \textbf{C}_i \textbf{C}_i^2
 f_i \rm{d} \textbf{c}_i
 \label{QKI}
\end{equation}

The above hydrodynamic equations (Eq. (\ref{numberconversation}), Eq. (\ref{momentconservation}), Eq. (\ref{energyconversation})) are based on the continuity hypothesis for the granular system. So  the established range of our results are limited to the Navier-Stokes level.

\section{\label{sec:level31}Constitutional relations}
In this section, we aim at deriving several terms, such as $\textbf{P}_{k,i}$ and $\textbf{q}_{k,i}$. In these derivations, the pair distribution function is employed as Eq. (\ref{pairdistirbution}).

To begin with, we calculate the stress tensor $\textbf{P}$ which is caused by two mechanisms, each contributing one part to $\textbf{P}$ \cite{Poschel}. The first is the kinetic part of the stress tensor, denoted by $\textbf{P}_{k,i}$ as defined in Eq. (\ref{PKI}). It results from the motion of all particles with mean velocity $\textbf{v}_{i}$ without any effect of particle-particle collisions.

Calculated from the first-order approximation to the distribution $f_{i}$, $\textbf{P}_{k,i}$ becomes

\begin{equation}
\textbf{P}_{k,i} =n_i \theta_i \hat{\textbf{I}} -  \frac{2 \mu_{i,dil}}{g_{ii}}
    \bigg(1+\frac{2\pi}{15} d^{3}_{i} n_{i}g_{ii}(1+e_{i}) \bigg)
    \mathring{\nabla}^{s}v_{i}
\label{pressurekinetic}
\end{equation}
where $\hat{\textbf{I}}$ is the unit tensor, $e_i$ is the restitution coefficient between particles in species $i$, and $\mu_{i,dil}$  is the viscosity for dilute suspensions \cite {Gidaspow1}, expressed as
\begin{equation}
  \mu_{i, dil}=\frac{5 m_i \sqrt{\pi\theta_{i}} }{16 \pi d_i^2}
\end{equation}

The second part of the stress tensor $\textbf{P}$ describes the momentum transfer caused by collisions. It is denoted by  $\textbf{P}_{c,ij}(m_i \textbf{C}_i)$. We obtain it by setting $\psi_{i}=m_i \textbf{C}_i$ in Eq. (\ref{generalPc}).
\begin{equation}
\begin{aligned}
\textbf{P}_{c,ij}(m_i \textbf{C}_i)
=&-\frac{1}{2}d_{ij}^3 \int_{\textbf{k} \cdot \textbf{c}_{ij}>0}
\left(m_i \textbf{C}_i' - m_i \textbf{C}_i\right)
\left(\textbf{k} \cdot \textbf{c}_{ij} \right)\textbf{k} \\&\times
f_{ij}^{(2)}\left(\textbf{c}_i,\textbf{r}-\frac{1}{2}d_{ij}\textbf{k},\textbf{c}_j,\textbf{r}+\frac{1}{2}d_{ij}\textbf{k}\right)
\rm{d} \textbf{k}\rm{d} \textbf{c}_i \rm{d} \textbf{c}_j
\end{aligned}
\end{equation}
 We still adopt the expansion of $f_{ij}^{(2)}\left(\textbf{c}_i,\textbf{r}-\frac{1}{2}d_{ij}\textbf{k},\textbf{c}_j,\textbf{r}+\frac{1}{2}d_{ij}\textbf{k}\right)$ in the paper \cite {PhysRevE.64.061301}, then,
\begin{equation}
\begin{aligned}
\label{pintegration}
\textbf{P}_{c,ij}(m_i \textbf{C}_i)
  =&\sum _{j} [-\frac{1}{2}g_{ij}d_{ij}^{3}\int _{\textbf{c}_{ij} \cdot \textbf{k}>0} m_{i}(\textbf{C}'_{i}-\textbf{C}_{i})(\textbf{c}_{ij}\cdot \textbf{k})\textbf{k}f_{i}f_{j}\rm{d} \textbf{k}\rm{d} \textbf{c}_i \rm{d} \textbf{c}_j  \\
  &-\frac{1}{4}g_{ij}d_{ij}^{4}\int _{\textbf{k} \cdot \textbf{c}_{ij}>0} m_{i}(\textbf{C}'_{i}-\textbf{C}_{i})(\textbf{c}_{ij}\cdot \textbf{k})\textbf{k}
  f_i f_j\nabla ln \frac{f_{j}}{f_{i}}\rm{d} \textbf{k}\rm{d} \textbf{c}_i \rm{d} \textbf{c}_j]   \\
  =&\sum_{j} (\textbf{P}_{c,ij}^{1}+ \textbf{P}_{c,ij}^{2})
\end{aligned}
\end{equation}
 where $ \textbf{P}_{c,ij}^{1} $ and $\textbf{P}_{c,ij}^{2}$ indicate above two integrations, separatively. We assume the diffusion velocity $\textbf{u}_i$ and $\textbf{u}_j$ approaches zero, so a factor $\exp(-\frac{m_i}{2\theta_i}\textbf{u}_i^2-\frac{m_j}{2\theta_j}\textbf{u}_j^2)$ becomes $1$. Using the integration technology \cite{Chapman,Ferziger}, we get
 \begin{equation}
 \begin{aligned}
\textbf{P}_{c,ij}^1= \frac{\pi}{48}\frac{1}{A^{3/2}D^{5/2}}d_{ij}^3 \varOmega
    \left(1 +\frac{5 B^2}{2 A D} + \frac{35 B^4}{8 A^2 D^2} + \ldots \right)\hat{\textbf{I}}
    \label{pressurecollision1}
\end{aligned}
\end{equation}
and
\begin{equation}
\begin{aligned}
\textbf{P}_{c,ij}^2 =
   \frac{\sqrt{\pi}}{96}d_{ij}^4 \varOmega
\bigg\{
   \left(B\frac{m_i}{\theta_i}R_{1}-\frac{1}{\theta_i}\frac{m_im_j}{m_0}R_{7} \right)
        \hat{\mathcal{S}}\left[\nabla,\textbf{v}_i\right] \\
   -\left(B\frac{m_j}{\theta_j}R_{1}+\frac{1}{\theta_j}\frac{m_im_j}{m_0}R_{7} \right)
        \hat{\mathcal{S}}\left[\nabla,\textbf{v}_j\right]
    \bigg\}
    \label{pressurecollision2}
\end{aligned}
\end{equation}
where we denote a tensor $\hat{\mathcal{S}}$ as the function of two vectors $\textbf{v}$ and $\textbf{w}$,
\begin{equation}
\hat{\mathcal{S}}[\textbf{v},\textbf{w}] = \frac{4}{5}[\mathring{\textbf{v}\textbf{w}}]^{s}+\frac{2}{3}[\textbf{v}\cdot\textbf{w}]\hat{\textbf{I}}
\end{equation}
$\textbf{v}\textbf{w}$ means the dyadic of two vectors. For any tensor $\hat{\textbf{X}}$, the superscript means

\begin{equation}
\mathring{\hat{\textbf{X}}}^{s}=\frac{1}{2}(\hat{\textbf{X}}+\overline{\hat{\textbf{X}} })- \frac{1}{3}\left(\hat{\textbf{X}}:\hat{\textbf{I}}\right)
\end{equation}
where the notation $\overline{\textbf{X}}$ is conjugate of $\textbf{X}$. And the following substitutions are used in Eq. (\ref{pressurecollision2}).
$$
R_{1}=\frac{1}{A^{5/2}D^3}, \;\,
R_{2}=\frac{1}{A^{5/2}D^{7/2}}, \;\,
R_{3}=\frac{1}{A^{5/2}D^4}, \;\,
R_{4}=\frac{1}{A^{7/2}D^3}, \;\,
R_{5}=\frac{1}{A^{7/2}D^4},
$$
$$
R_{6}\!=\!\frac{1}{A^{3/2}D^{5/2}}\!\left(\!1\!+\!\frac{5B^2}{2AD}\right)\!, \quad
R_{7}\!=\!\frac{1}{A^{3/2}D^3}\!\left(\!1\!+\!\frac{3B^2}{AD}\right)\!,        \,
R_{8}\!=\!\frac{1}{A^{3/2}D^{7/2}}\!\left(\!1\!+\!\frac{7B^2}{2AD}\right)\!,
$$
$$
R_9\!=\!\frac{1}{A^{5/2}D^{5/2}}\!\left(\!1\!+\!\frac{25B^2}{6AD}\right)\!,
R_{10}\!=\!\frac{1}{A^{5/2}D^{5/2}}\!\left(\!1\!+\!\frac{11B^2}{2AD}\right)\!,
R_{11}\!=\!\frac{1}{A^{5/2}D^3}\!\left(\!1\!+\!\frac{5B^2}{AD}\right)\!,
$$
$$
R_{12}=\frac{1}{A^{5/2}D^4}\left(1+\frac{4B^2}{AD}\right),
\varOmega=(1+e_{ij})\frac{m_im_j}{m_0} g_{ij} n_i n_j
    \left(\frac{m_i m_j}{\theta _i \theta _j}\right)^{3/2} \qquad \qquad \qquad
$$
Consequently, we acquire the total particle pressure $\textbf{p}_{i}$ from the normal term of the sum of $\textbf{P}_{k,i}$ (Eq. (\ref{pressurekinetic})) and $\textbf{P}_{c,ij}$ (Eq. (\ref{pressurecollision1}), Eq. (\ref{pressurecollision2}))

\begin{equation}
\begin{aligned}
\textbf{p}_{i}=&n_{i} \theta_{i}
    +\sum_{j}
\frac{\pi}{48}
 \frac{1}{A^{3/2} D^{5/2}} d_{i j}^3 \varOmega
 \left( 1 + \frac{5 B^2}{2 A D} + \frac{35 B^4}{8 A^2 D^2} + \ldots \right)
\end{aligned}
\end{equation}

In addition, the coefficient of viscosity $\mu_i$ is defined from the the shear term of the sum of $\textbf{P}_{k,i}$ (Eq. (\ref{pressurekinetic})) and $\textbf{P}_{c,ij}$ (Eq. (\ref{pressurecollision1}), Eq. (\ref{pressurecollision2})), thus

\begin{equation}
\label{viscosity}
\begin{aligned}
\mu_{i}=&
    \frac{\mu_{i,dil} }{g_{ii}}
       \bigg(1+\frac{2\pi}{15} d^{3}_{i} n_{i}g_{ii}(1+e_{i}) \bigg)
        \bigg(1+\frac{2\pi}{15} d^{3}_{i} n_{i}g_{ii}(1+e_{i})
             \\  +& \frac{2\pi}{15} \frac{n_jm_j}{m_0}d^{3}_{ij}g_{ij}(1+e_{ij}) \bigg)
        + \sum_{j}
    { \frac{\sqrt{\pi}}{240}
     d_{i j}^4  \varOmega
     \left( \frac{m_im_j}{ m_0} \frac{1}{\theta _i } R_{7} -
     B \frac{m_i}{\theta _i} R_{1} \right)}
\end{aligned}
\end{equation}

The momentum source term $\psi_i$ involved in Eq. (\ref{momentconservation}) is therefore
\begin{equation}
\begin{aligned}
&\textbf{N}_{ij}(m_i \textbf{C}_i) \\
& =d_{ij}^2 \int_{\textbf{k} \cdot \textbf{c}_{ij}>0}
\left(m_i \textbf{C}_i' - m_i \textbf{C}_i\right)
\left(\textbf{k} \cdot \textbf{c}_{ij} \right)
f_{ij}^{(2)}\left(\textbf{c}_i,\textbf{r}-\frac{1}{2}d_{ij}\textbf{k},\textbf{c}_j,\textbf{r}+\frac{1}{2}d_{ij}\textbf{k}\right)
\rm{d} \textbf{k}\rm{d} \textbf{c}_i \rm{d} \textbf{c}_j \\
&=\sum _{j} \phi_{ij}^1 + \phi_{ij}^2
\end{aligned}
\end{equation}
where
\begin{equation}
\phi_{ij}^1=-\frac{m_{i}m_{j}}{m_{0}}d^{2}_{ij}(1+e_{ij})\int_{ \textbf{c}_{ij} \cdot \textbf{k} >0} (\textbf{c}_{ij} \cdot \textbf{k})^2 \textbf{k}f_{i}f_{j} \rm{d} \textbf{k}\rm{d} \textbf{c}_i \rm{d} \textbf{c}_j
\end{equation}
\begin{equation}
 \phi_{ij}^2=-\frac{m_{i}m_{j}}{m_{0}}\frac{d^{3}_{ij}}{2}(1+e_{ij})\int_{ \textbf{c}_{ij} \cdot \textbf{k} >0} (\textbf{c}_{ij} \cdot \textbf{k})^2 \textbf{k}f_{i}f_{j}\textbf{k} \cdot \nabla \ln \frac{f_{j}}{f_{i}}  \rm{d} \textbf{k}\rm{d} \textbf{c}_i \rm{d} \textbf{c}_j
\end{equation}
whence, to the same approximation and integration,

\begin{equation}
\begin{aligned}
\phi_{ij}^1 &=
    \frac{1}{12} \frac{\sqrt{\pi}}{A^{5/2} D^{3}} d_{ij}^2 \varOmega
    \bigg[ \frac{m_i m_j}{m_0}\left(\frac{\textbf{u}_j}{\theta
   _j}-\frac{\textbf{u}_i}{\theta_i}\right) \! \left(A+\frac{3 B^2}{D}\right)+
    B\left(\frac{m_i }{\theta_i}\textbf{u}_i+\frac{m_j }{\theta_j}\textbf{u}_j\right) \!
\bigg]
\end{aligned}
\end{equation}

\begin{equation}
\begin{aligned}
\phi_{ij}^2 &=
    \frac{\pi}{32}
    d_{ij}^3 \varOmega
    \bigg \{ \nabla \ln \frac{\theta_j}{\theta_i} R_{6}
    -\frac{1}{2} \left(\frac{m_j \nabla \theta_j}{\theta_j^2} - \frac{m_i \nabla \theta_i}{\theta_i^2} \right) R_9 \\ &
    -\frac{5}{6}\frac{m_i m_j}{m_0^2} \left(\frac{m_i \nabla \theta_j}{\theta_j^2} - \frac{m_j \nabla \theta_i}{\theta_i^2} \right)
        R_{8}
    - \frac{5B}{3}\frac{m_im_j}{m_0}\left(\frac{\nabla \theta_j}{\theta_j^2}+\frac{\nabla \theta_i}{\theta_i^2} \right)R_{2}
    \bigg \}
\end{aligned}
\end{equation}

As defined in Eq. (\ref{QKI}), $ \textbf{q}_{k,i}$ denotes the kinetic energy flux due to the particle transport without collisions.
Using the zeroth-order approximation, $ \textbf{q}_{k,i}$ is found to be equal to zero, since the integration is on the odd power of the velocity $\textbf{C}_i$.

Extending $ \textbf{q}_{k,i}$ to the first-order approximation, we obtain
 \begin{equation}
   \textbf{q}_{k,i}^{(1)} = \frac{\kappa_{i,dil}}{g_{ii}} \left[ 1 + \frac{\pi}{10} n_{i} d^{3}_{i} g_{ii}(1+e_{i})^2 \right] \nabla \theta_i
 \end{equation}
where $\kappa_{i,dil}$ is the conductivity for dilute particle phase, expressed as $\kappa_{i,dil}=\frac{5}{2}\mu_{i,dil}C_{V}$, and $ C_{V}=\frac{3}{2m_i}$.

The collisional contribution to the energy flux is determined by
$\textbf{P}_{ij} \left(  \frac{1}{2}m_i \textbf{C}_i^2 \right)$, where
\begin{equation}
 \begin{aligned}
 &\textbf{P}_{ij} \left(  \frac{1}{2}m_i \textbf{C}_i^2 \right)  \\
 &=-\frac{m_i}{4}d_{ij}^3 \int_{\textbf{k} \cdot \textbf{c}_{ij}>0} \!\!
 \left( \textbf{C}_i'^2- \textbf{C}_i^2 \right)
 \left(\textbf{k} \cdot \textbf{c}_{ij} \right)\textbf{k}
 f_{ij}^{(2)}\left(\textbf{c}_i,\textbf{r}-\frac{d_{ij}}{2}\textbf{k},\textbf{c}_j,\textbf{r}+\frac{d_{ij}}{2}\textbf{k}\right)
\rm{d} \textbf{k}\rm{d} \textbf{c}_i \rm{d} \textbf{c}_j      \\
 &=\textbf{q}_{c,ij}^1 + \textbf{q}_{c,ij}^2
 \end{aligned}
\end{equation}
and
\begin{equation}
  \textbf{q}_{c,ij}^{1} =-\frac{d_{ij}^{3}}{2} \int_{\textbf{k} \cdot\textbf{ c}_{ij}>0}
 \left( \frac{1}{2}m_i \textbf{C}_i'^2-\frac{1}{2}m_i \textbf{C}_i^2 \right)
 \left(\textbf{k} \cdot \textbf{c}_{ij} \right)\textbf{k} f_{i} f_{j}
\rm{d} \textbf{k}\rm{d} \textbf{c}_i \rm{d} \textbf{c}_j
  \label{q1}
\end{equation}
\begin{equation}
   \textbf{q}_{c,ij}^{2} =-\frac{d_{ij}^{4}}{4} \int_{\textbf{k} \cdot \textbf{c}_{ij}>0}
 \left( \frac{1}{2}m_i \textbf{C}_i'^2-\frac{1}{2}m_i \textbf{C}_i^2 \right)
 \left(\textbf{k} \cdot \textbf{c}_{ij} \right)\textbf{k} f_{i} f_{j} \textbf{k} \cdot \nabla\ln \frac{f_{j}}{f_{i}}
\rm{d} \textbf{k}\rm{d} \textbf{c}_i \rm{d} \textbf{c}_j
    \label{q2}
\end{equation}
Among them, $\frac{1}{2}m_i(\textbf{C}_{i}'^{2}- \textbf{C}_{i}^2)$ is the kinetic energy change for species $i$ in one collision. As noted above, the collisions between the particles are inelastic, so the kinetic energy change during a collision is
\begin{equation}
\frac{1}{2}m_i \textbf{C}_i'^2-\frac{1}{2}m_i \textbf{C}_i^2 =
\frac{m_i}{2}\left[\frac{m_j^2}{m_0^2}(1+e_{ij})^2(\textbf{c}_{ij} \cdot \textbf{k})^2 -2
\frac{m_j}{m_0}(1+e_{ij})(\textbf{c}_{ij} \cdot \textbf{k})(\textbf{C}_{i} \cdot \textbf{k})
\right]
\label{collsionloss}
\end{equation}
Substituting Eq. (\ref{collsionloss}) into Eq. (\ref{q1}) and Eq. (\ref{q2}), we obtain
\begin{equation}
\begin{aligned}
 \textbf{q}_{c,ij}^1 =  &\frac{\pi}{96}  d_{ij}^3 \varOmega
 \bigg \{
    \left( \frac{m_i \textbf{u}_i}{\theta_i} + \frac{m_j \textbf{u}_j}{\theta_j} \right)
    \left( R_{10}+\frac{3B}{2}\frac{m_j}{m_0}(e_{ij}-1)R_{2} \right)  \\
   -&\frac{m_im_j}{m_0}\left( \frac{\textbf{u}_i}{\theta_i} -\frac{\textbf{u}_j}{\theta_j}\right)
   \left(3BR_{2}+\frac{3}{2}\frac{m_j}{m_0}(e_{ij}-1)R_{8}\right)
 \bigg \}
\end{aligned}
\end{equation}
\begin{equation}
 \begin{aligned}
\textbf{q}_{c,ij}^2
  &=\frac{\sqrt{\pi}}{384}
   d_{i j}^4 \varOmega
   \bigg\{
   -\frac{4m_i m_j }{m_0}\left[R_{1} + 9B^2R_{5}
     + 3B \frac{m_j}{m_0}(e_{ij}-1)R_{3} \right]
       \left(\frac{\nabla \theta_i}{\theta_i^2} + \frac{\nabla \theta_j}{\theta_j^2}\right)          \\
   &+ \left[ 6\frac{m_j}{m_0}(e_{ij}-1)R_{7} + 12BR_{1}\right]
     \nabla \ln \frac{\theta_j}{\theta_i}
   - \left[ \frac{3m_j}{m_0}(e_{ij}-1)R_{11} + 10 B R_{4} \right]                       \\
   &\times  \left(\frac{m_j \nabla \theta_j}{\theta_j^2}-\frac{m_i \nabla \theta_i}{\theta_i^2}\right)
   - \frac{6m_im_j}{m_0^2}
     \left[ \frac{m_j}{m_0}(e_{ij}-1)R_{12} + 2 B R_{3} \right]
     \left( \frac{m_i \nabla \theta_j}{\theta_j^2} - \frac{m_j \nabla \theta_i}{\theta_i^2} \right)  \\
   &-
     \frac{2m_j}{ m_0}\left(e_{ij}-1\right)
     \bigg[
     \frac{m_j}{\theta_j} \left(\frac{m_i}{m_0}R_{7}+BR_{1}\right)
    \hat{\mathcal{S}}[\nabla,\textbf{v}_j]  \cdot \textbf{u}_i                          \\
   &+
    \frac{m_i}{\theta_i} \left(\frac{m_j}{m_0}R_{7}-BR_{1}\right)
    \hat{\mathcal{S}}[\nabla,\textbf{v}_i] \cdot \textbf{u}_i
     \bigg]
     \bigg\}
 \end{aligned}
\end{equation}

Finally, we substitute $\psi_i$ with energy $\frac{1}{2}m_i \textbf{C}_i^2$, then collision part of energy dissipation term can be written as
\begin{equation}
 \begin{aligned}
 &\textbf{N}_{ij} \left(\frac{1}{2}m_i \textbf{C}_i^2\right) \\
 &=d_{ij}^2 \int_{\textbf{k} \cdot \textbf{c}_{ij}>0}
  \frac{m_i}{2}\left(  \textbf{C}_i'^2- \textbf{C}_i^2 \right)
  \left(\textbf{k} \cdot \textbf{c}_{ij} \right)
  f_{ij}^{(2)}\left(\textbf{c}_i,\textbf{r}-\frac{1}{2}d_{ij}\textbf{k},\textbf{c}_j,\textbf{r}+\frac{1}{2}d_{ij}\textbf{k}\right)
  \rm{d} \textbf{k}\rm{d} \textbf{c}_i \rm{d} \textbf{c}_j  \\
 &=\gamma_{c,ij}^1+\gamma_{c,ij}^2
  \end{aligned}
\end{equation}
where
\begin{equation}
 \begin{aligned}
 \gamma_{c,ij}^1 = d_{ij}^2 g_{ij}\int_{k \cdot c_{ij}>0}
  \left( \frac{1}{2}m_i \textbf{C}_i'^2-\frac{1}{2}m_i \textbf{C}_i^2 \right)
  \left(\textbf{k} \cdot \textbf{c}_{ij} \right)
  f_i f_j
  \rm{d} \textbf{k}\rm{d} \textbf{c}_i \rm{d} \textbf{c}_j
  \end{aligned}
\end{equation}
\begin{equation}
 \begin{aligned}
  \gamma_{c,ij}^2 =\frac{1}{2} d_{ij}^3 g_{ij} \int_{k \cdot c_{ij}>0}
  \left( \frac{1}{2}m_i \textbf{C}_i'^2-\frac{1}{2}m_i \textbf{C}_i^2 \right)
  \left(\textbf{k} \cdot \textbf{c}_{ij} \right)
  f_{i} f_{j}\nabla \ln \frac{f_j}{f_i}
  \rm{d} \textbf{k}\rm{d} \textbf{c}_i \rm{d} \textbf{c}_j
  \end{aligned}
\end{equation}
Following the above integration method, $\gamma_{c,ij}$ includes
\begin{equation}
\begin{aligned}
 \gamma_{c,ij}^1 &= \frac{\sqrt{\pi}}{8} \frac{1}{A^{3/2}D^3}
    d_{ij}^2 \varOmega
  \bigg\{
    \left(1+\frac{3B^2}{AD}\right)
    \bigg[
        \frac{m_j}{m_0}(e_{ij}-1) \\ &+
        \frac{2}{3} \frac{m_i m_j}{m_0}\textbf{u}_i \cdot
        \left(\frac{\textbf{u}_i}{\theta_i} - \frac{\textbf{u}_j}{\theta_j} \right)
    \bigg]
    +\frac{2}{3}\frac{B}{A}
    \left[3 - \textbf{u}_i\cdot \left( \frac{m_i}{\theta_i} \textbf{u}_i+\frac{m_j}{\theta_j} \textbf{u}_j \right)
    \right]
 \bigg\}
\end{aligned}
\label{gamma1}
\end{equation}

and
\begin{equation}
 \begin{aligned}
  \gamma_{c,ij}^2  =&  \frac{\pi}{16}
    d_{ij}^3 \varOmega
    \bigg \{
      \left[ \frac{m_j}{\theta_j}(\nabla\cdot \textbf{v}_{j})
         -\frac{m_i}{\theta_i}(\nabla\cdot \textbf{v}_{i}) \right]
      \left[ \frac{B}{4} \frac{m_j}{m_0}(1-e_{ij})  R_{2} +  \frac{1}{3}R_{10} \right] \\+&
      \frac{m_im_j}{m_0}
      \left(\frac{\nabla\cdot \textbf{v}_{j}}{\theta_j}+\frac{\nabla\cdot \textbf{v}_{i}}{\theta_i} \right)
      \left[ \frac{1}{4}\frac{m_j}{m_0}(1-e_{ij})R_{8} + B R_{2} \right]
    \bigg \}
 \end{aligned}
\end{equation}
\section{\label{sec:level41}Results and discussion}
We now evaluate our theory predictions by comparing with the previous works. In order to particularize the general results given in the the above section, we adopt the typical fluidized bed conditions used in Ref. \cite{ISI:000187361700002}. The parameters include masses $m_{i}=2.79\times10^{-10}~kg$, $m_{j}=6.62\times10^{-10}~kg$, diameters $d_{i}=d_{j}=5\times10^{-4}~m$, the restitution coefficients $e_{ij}=e_{i}=e_{j}=0.9$, the granular temperature ranges from $1.5\times10^{-11}~kg~m^{2}/s^{2}$ to $10\times10^{-11} ~kg~m^{2}/s^{2}$, the volume fractions $\varepsilon_{i}=\varepsilon_{j}=0.25$, $\varepsilon_{max}=0.638$.

\begin{figure}
\centering
 \includegraphics[width=0.8\textwidth]{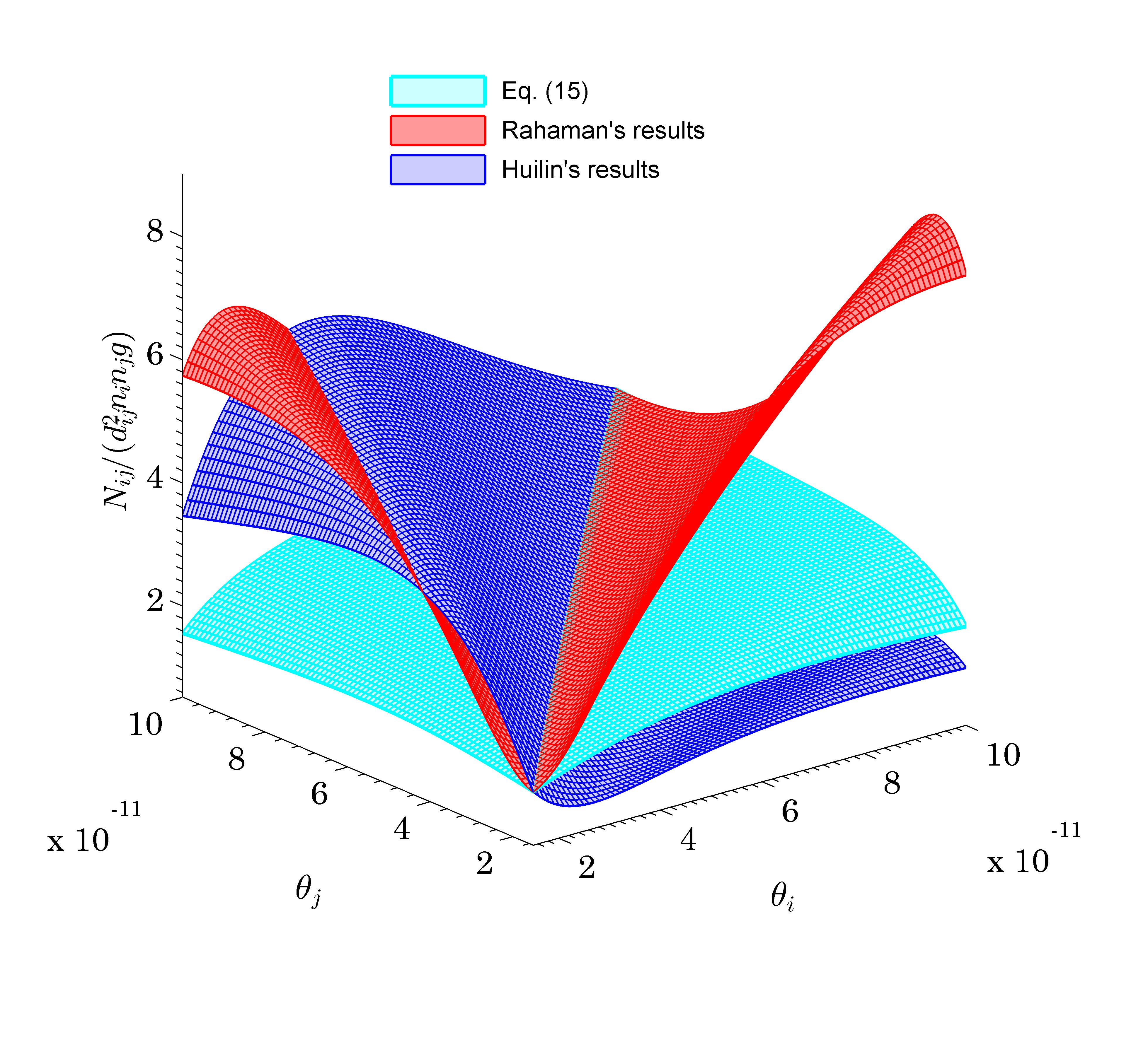}%
 \caption{Surfaces of scaled collision frequency in terms of two granular temperatures $\theta_{i}$ and $\theta_{j}$ of our results (Eq. (\ref{collisionfreqresult})), Rahaman's result \cite{ISI:000187361700002} and Lu's result \cite{PhysRevE.64.061301}, where $m_{i}=2.79\times10^{-10}~kg$, $m_{j}=6.62\times10^{-10}~kg$. $N_{ij}$ is scaled by $d^{2}_{ij}n_{i}n_{j}g$.}
 \label{collsionfre}
 \end{figure}

We begin with the collision frequency  $N_{ij}$ which presents the number of collisions between particles of species $i$ and $j$ per unit time and per unit volume. $N_{ij}$ is expected to be equal to $N_{ji}$ due to the symmetric character in collisions. In the $N_{ij}$ expression, the parameters $A$ and $D$ are symmetric about $i$ and $j$, but $B$ is not. To ensure symmetry of $N_{ij}$, the expression of $N_{ij}$ should only contain even powers of $B$. Our result ( Eq. (\ref{collisionfreqresult})) agrees well with this. Fig. (\ref{collsionfre}) plot $3D$ surfaces of scaled collision frequency of Eq. (\ref{collisionfreqresult}) in terms of granular temperature of two species and comparison with previous results \cite{PhysRevE.64.061301,ISI:000187361700002}.  $N_{ij}$ is scaled by $d^{2}_{ij}n_{i}n_{j}g$ considering that $N_{ij}$ vary greatly in different value of the Enskog factors (here, denoted by $g$). Three surfaces intersect at line $\theta_{i}=\theta_{j}$. Furthermore, we can observe that our surfaces are higher than the Lu's prediction and lower than Rahanman's when $\theta_{i}>\theta_{j}$. This is expected as the Lu's profile can only be applied in energy equipartition systems as pointed in \cite{ISI:000229306300005}, which may underestimate the collision frequencies, whereas Rahaman's hypothesis implies all the collisions happen in a plane, which obviously overestimates the collision frequency. When $\theta_{i}<\theta_{j}$, The value of Eq. (\ref{collisionfreqresult}) is still smaller than Rahanman's results. Variation or error of $N_{ij}$ of Lu's results become large due to negative B.

 \begin{figure}
\centering
 \includegraphics[width=0.8\textwidth]{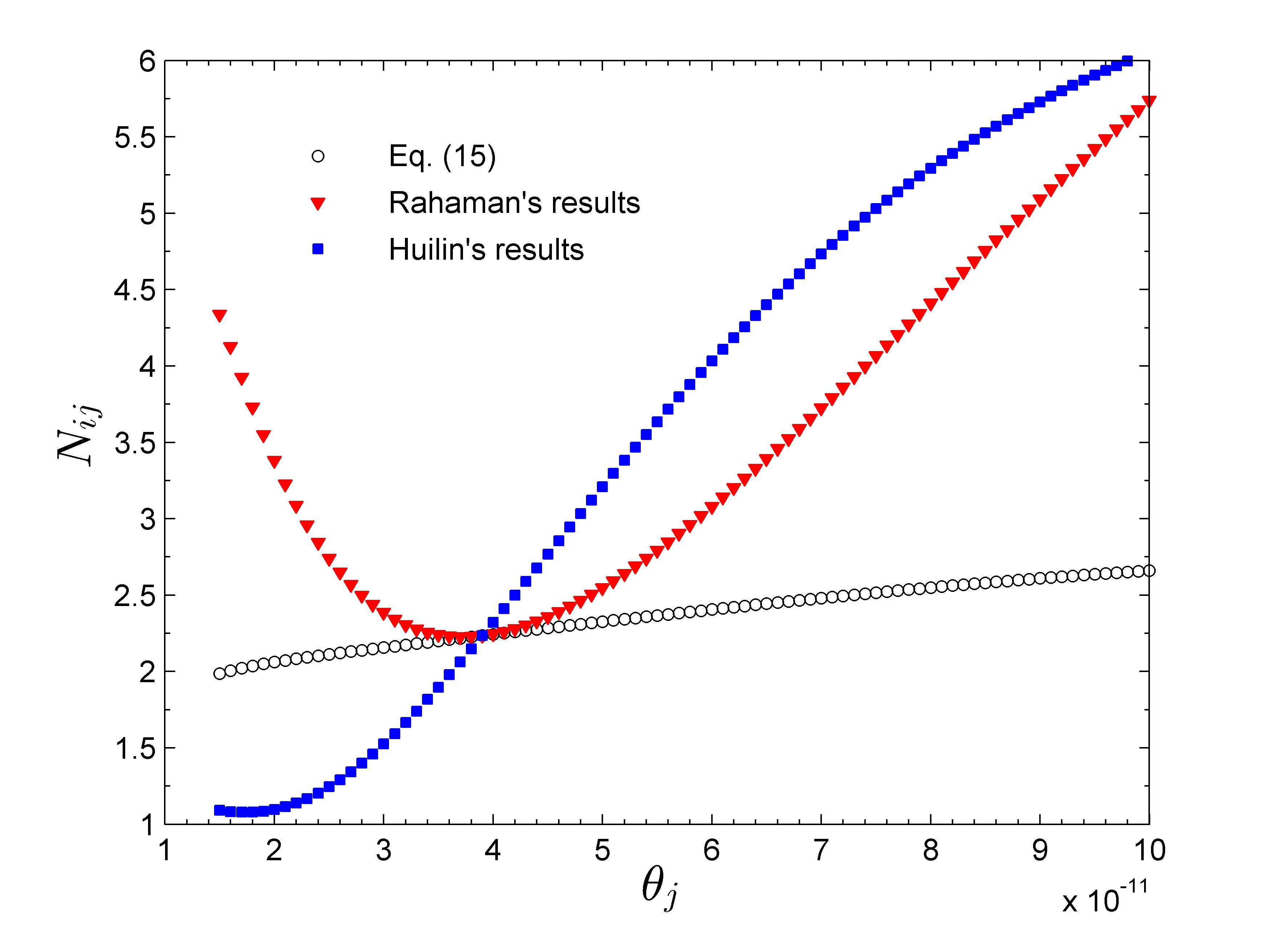}%
 \caption{Profiles of scaled collision frequency as a function of granular temperature $\theta_{j}$, where $\theta_{i}=4\times10^{-11}~kg~m^{2}/s^{2}$ , $m_{i}=2.79\times10^{-10}~kg$, and $m_{j}=6.62\times10^{-10}~kg$. $N_{ij}$ is scaled by $d^{2}_{ij}n_{i}n_{j}g$.}\label{linear2DNij}
 \end{figure}

To have a closer examination, we plot a cross section of Fig. (\ref{collsionfre}) as a function of granular temperature $\theta_{j}$ when $\theta_{i}=4\times10^{-11}~kg~m^{2}/s^{2}$  which is illustrated in Fig. (\ref{linear2DNij}). It can be seen that our results intersect with the Rahaman's and Lu's results at the point $\theta_{i}=\theta_{j}=4\times10^{-11}~kg~m^{2}/s^{2}$ which confirms that these three results are equal at $\theta_{i}=\theta_{j}$. Besides, we could find that Rahaman's $N_{ij}$ decreases in the region $\theta_{i}>\theta_{j}$, and increases in the region $\theta_{i}<\theta_{j}$ with increasing $\theta_{j}$, but our results show that $N_{ij}$ increases with increasing $\theta_{j}$ in both of two regions. As the collision frequency increases with the increase of granular temperature, the value of $N_{ij}$ of the region $\theta_{j}<\theta_{i}=4\times10^{-11}~kg~m^{2}/s^{2}$ is supposed to be smaller than that of point $\theta_{j}=\theta_{i}=4\times10^{-11}~kg~m^{2}/s^{2}$. So we can expect that our results are more reasonable than Rahaman's.

\begin{figure}
\centering
 \includegraphics[width=0.8\textwidth]{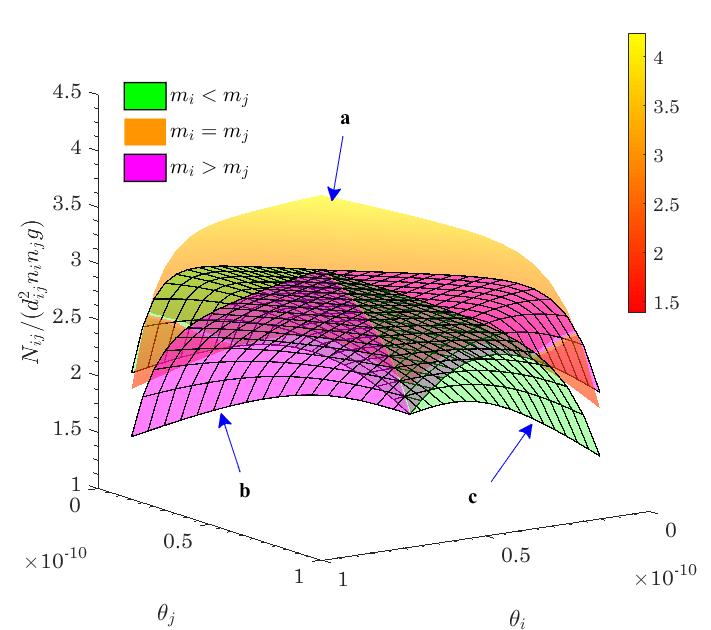}%
 \caption{Surfaces of scaled collision frequency in terms of two granular temperatures $\theta_{i}$ and $\theta_{j}$, where (a) $m_{i}=m_{j}=2.79\times10^{-10}~kg$; (b) $m_{j}=2.79\times10^{-10}~kg$, $m_{i}=6.62\times10^{-10}~kg$; (c) $m_{i}=2.79\times10^{-10}~kg$, $m_{j}=6.62\times10^{-10}~kg$. $N_{ij}$ is scaled by $d^{2}_{ij}n_{i}n_{j}g$.}\label{masscompare}
 \end{figure}

In Fig. (\ref{masscompare}), We plot $N_{ij}$ in Eq. (\ref{collisionfreqresult}) in terms of granular temperature of two species under various mass ratios, (a) $m_{i}=m_{j}=2.79\times10^{-10}~kg$; (b) $m_{i}=2.79\times10^{-10}~kg$, $m_{j}=6.62\times10^{-10}~kg$; (c) $m_{j}=2.79\times10^{-10}~kg$, $m_{i}=6.62\times10^{-10}~kg$. Surface of collision frequency $N_{ij}$ with $m_{i}=m_{j}$ (a) is symmetric about the $\theta_{i} =\theta_{j}$.  If $m_{i} \neq m_{j}$, the surface of collision frequency $N_{ij}$ tilts about $\theta_{i} =\theta_{j}$ axis. We exchanged the mass $m_{i}$ and $m_{j}$, the surface (b) and (c) is symmetric about $\theta_{i} =\theta_{j}$.
%
\begin{figure}
\centering
 \includegraphics[width=0.8\textwidth]{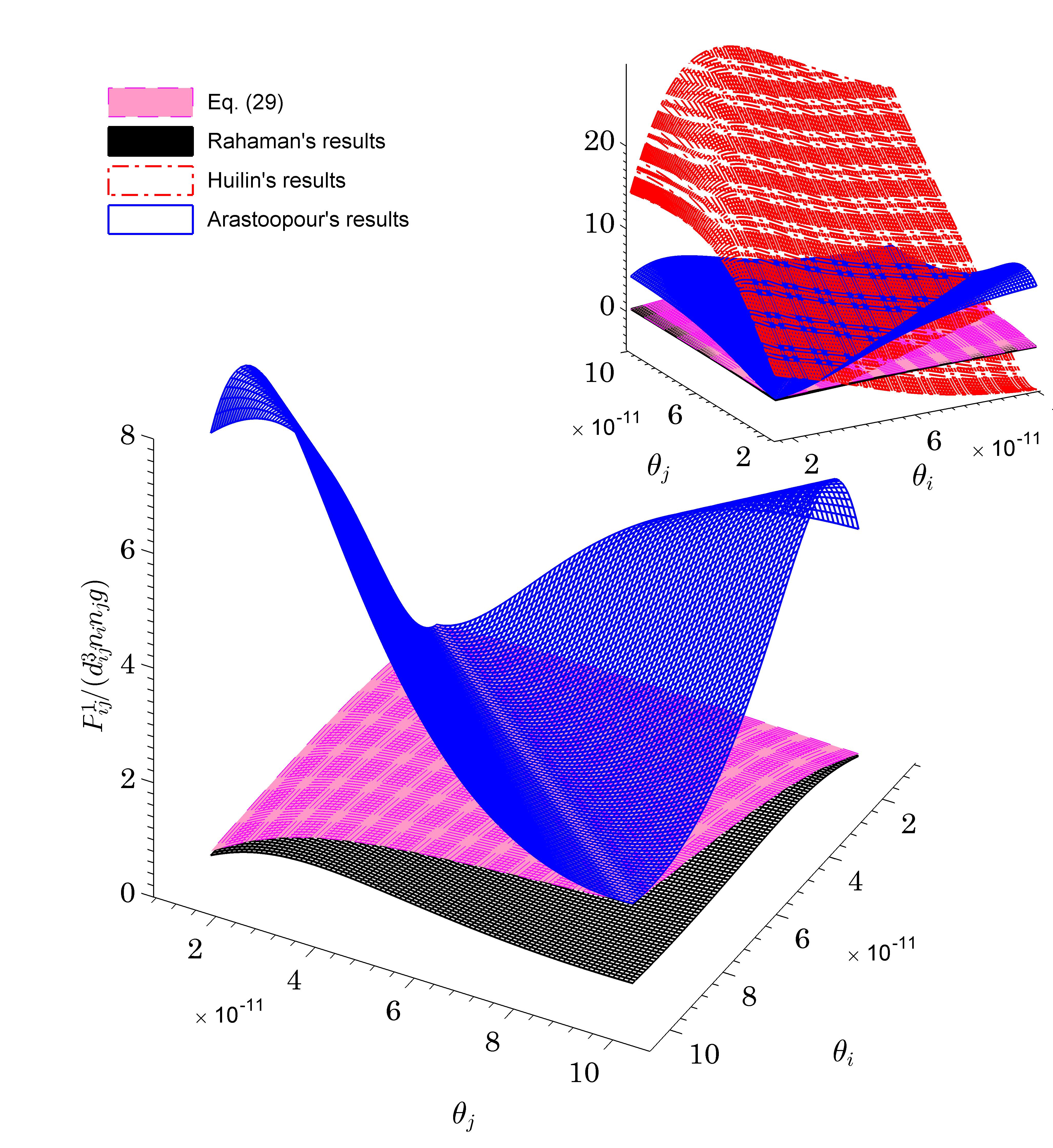}%
 \caption{Surfaces of collision stress $\textbf{P}_{c,ij}^{1}$ as a function of granular temperature $\theta_{i}$ and $\theta_{j}$, where $m_{i}=2.79\times10^{-10}~kg$, $m_{j}=6.62\times10^{-10}~kg$, and $e_{ij}=0.9$. The main figure includes three results: Rahaman's, Arastoopour's and Eq. (\ref{pintegration}) and the inset figure adds Lu's results.  }
 \label{mixturespressurecompar}
\end{figure}
Fig. (\ref{mixturespressurecompar}) show the variation of the collisional stress component (zeroth-order approximation) with two granular temperatures. The main figure includes three results: Rahaman's, Arastoopour's and Eq. (\ref{pintegration}), and the inset adds Lu's results. It could be found that our results are tangent to Arastoopour's at $\theta_{i}=\theta_{j}$ and are very close to Rahaman's results. In Arastoopour's results \cite{ISI:000229306300005}, $\textbf{P}_{c,ij}^{1}$ reaches its maximum near the point of maximum of  $|\theta_{i}-\theta_{j}|$, i.e $(1.5\times10^{-11}~kg~m^{2}/s^{2},10\times10^{-11}~kg~m^{2}/s^{2})$ and $(10\times10^{-11}~kg~m^{2}/s^{2},1.5\times10^{-11}~kg~m^{2}/s^{2})$ in Fig. (\ref{mixturespressurecompar}). While our results shows the maximum of $\textbf{P}_{c,ij}^{1}$ at maximum of both of $\theta_{i}$ and $\theta_{j}$, i.e. $(10\times10^{-11}~kg~m^{2}/s^{2},10\times10^{-11}~kg~m^{2}/s^{2})$ in Fig. (\ref{mixturespressurecompar}). Such difference may be caused by that Arastoopour's work \cite{ISI:000229306300005} is based on the assumption that interaction between particles from two species is only at the interface, while our results are free of this assumption and hence the integration of Eq. (\ref{pintegration}) is complete.

\begin{figure}
\centering
 \includegraphics[width=0.8\textwidth]{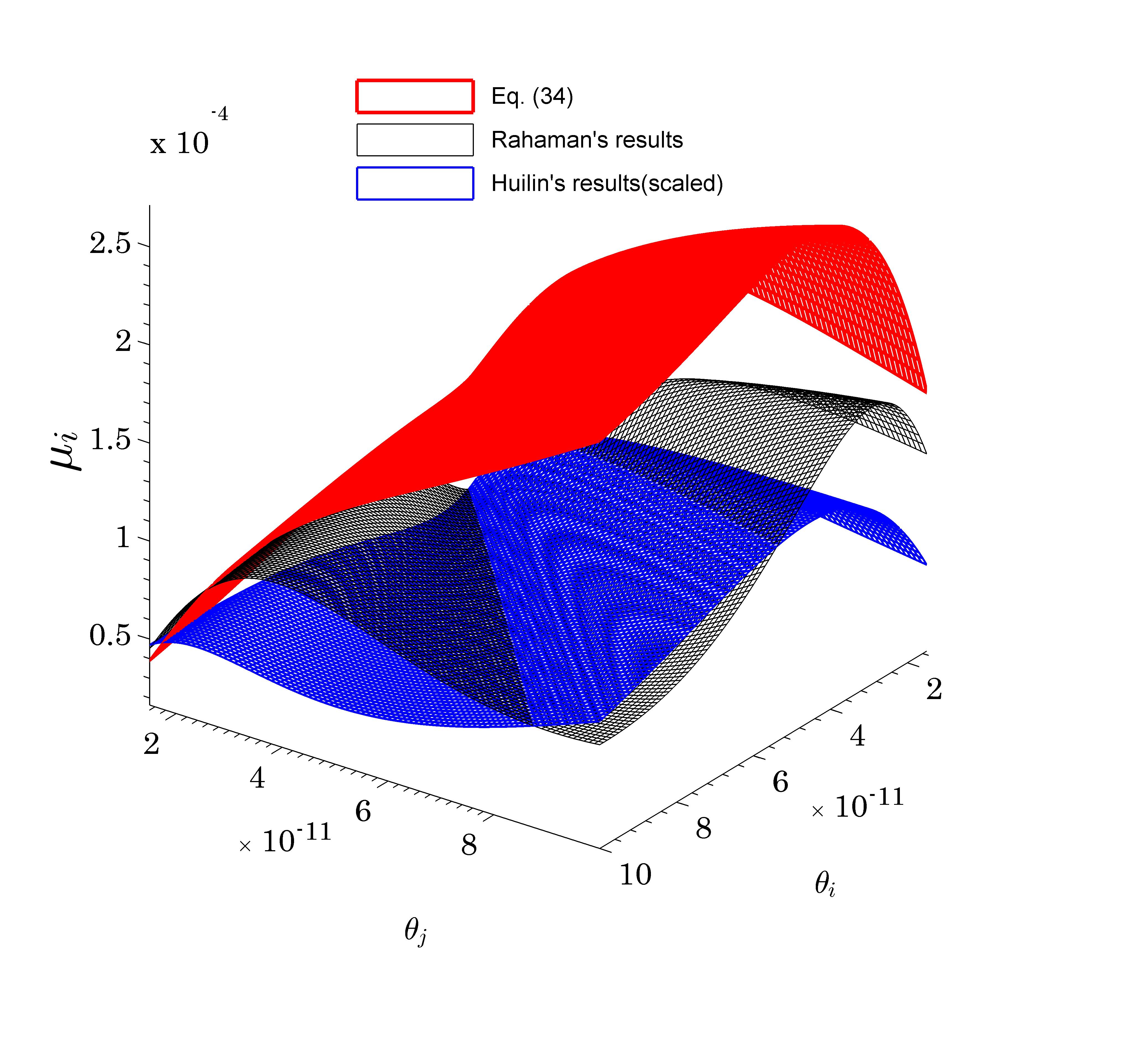}%
 \caption{Surfaces of scaled viscosity as a function of granular temperature $\theta_{j}$ and $\theta_{j}$,  where $m_{i}=2.79\times10^{-10}~kg$, $m_{j}=6.62\times10^{-10}~kg$, $\varepsilon_{i}=\varepsilon_{j}=0.25$, $\varepsilon_{max}=0.638$, $d_{i}=d_{j}=5\times10^{-4}~m$, and $e_i=e_{ij}=0.9$.}\label{viscosity_fig}
 \end{figure}

 Fig. (\ref{viscosity_fig}) plots the solids shear viscosity (Eq. (\ref{viscosity})) and its comparison with Lu's \cite{PhysRevE.64.061301} and Rahman's \cite{ISI:000187361700002} results. Because the Enskog factor in Lu's paper is much larger than the rest ($g=58$ in paper \cite{PhysRevE.64.061301}, $g=1.68$ in paper \cite{ISI:000187361700002} and $g=5.52$ in our case (Eq. (\ref{genskog})) ), we scaled Lu's results \cite{PhysRevE.64.061301} to make these three viscosities can be drawn in one figure. Unlike Rahaman's result where the viscosity has a minimum at $\theta_{i}=\theta_{j}$, our result shows that the $\mu_{i}$ is still mainly determined by $\theta_{i}$ and the maximum of viscosity appears at the maximum of $\theta_{i}$, which demonstrates that the dilute viscosity increases with the increase of the granular temperature.

\begin{figure}
\centering
 \includegraphics[width=0.8\textwidth]{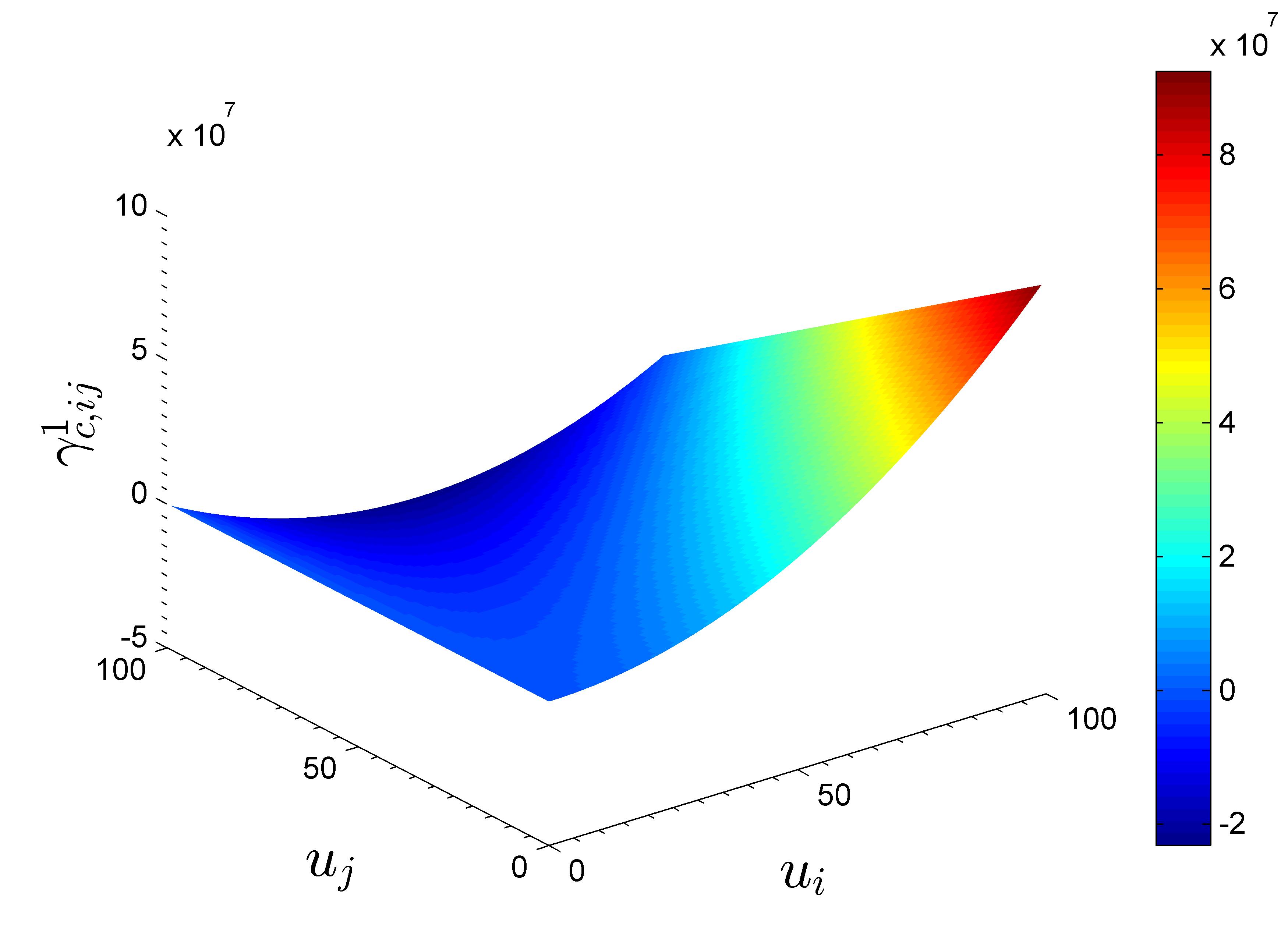}%
 \caption{Surfaces of the energy dissipation $\gamma _{c,ij}^{1}$ as a function of mean velocities $\textbf{u}_{i}$ and $\textbf{u}_{j}$,  where $\varepsilon_{i}=\varepsilon_{j}=0.25$, $\varepsilon_{max}=0.638$, $d_{i}=d_{j}=5\times10^{-4}~m$, $\theta_{i}=\theta_{j}= 3.9000\times10^{-11} kg ~ m^{2}/s^{2}$, $m_{i}=2.79\times10^{-10}~kg$, $m_{j}=6.62\times10^{-10}~kg$ and $e_{ij}=0.9$.}\label{energysource}
 \end{figure}

The model proposed in this article allows different granular temperatures, mean velocities and diameters, masses, etc. In the previous discussion, the parameters such as the collision frequency and shear viscosity seem to be only affected by granular temperatures and masses, but not by mean velocities of particles of two species. Previous researches \cite{PhysRevE.64.061301,ISI:000187361700002} are not accountable to two different mean velocities, either. To investigate the influence of mean velocities, Fig. (\ref{energysource}) illustrates the surface of energy dissipation component $\gamma_{c,ij}^{1}$ (Eq. (\ref{gamma1})) in terms of two mean velocities. It can be found that when the mean velocities are not equal, $\gamma_{c,ij}^{1}$ varies greatly, reaching its peak when $\textbf{u}_{i}$ is maximum and $\textbf{u}_{j}$ is minimum. So the difference of two mean velocities should not be ignored. In an air-fluidized bed, the mean velocities of clusters and dispersed particles are usually not equal to each other. Our model is sensitive to this factor. And more elaborate validation needs further efforts.

\section{\label{sec:level61}Conclusion}

In this paper, we develop a kinetic theory based model for a three-dimensional binary granular mixture with different masses, sizes, mean velocities, densities as well as granular temperatures, using standard Enskog theory. The integration of the inner product of relative velocity $\textbf{c}_{ij}$ and combined velocity $\textbf{G}$ is properly treated in three-dimensional space.

The computed collision frequency and particle viscosity coefficient increase monotonically with the increase of granular temperature of each particle species. Our results also show that the energy dissipation component depends heavily on the mean velocities of particle species, thus, difference of mean velocities, if any, should not be overlooked.

Compared with previous work, our research enriches and consummates the previous theories and is more suitable for the multi-type particle theory in gas-solid flow systems. This work can be applied in the case that the effect of energy non-equipartition and unequal mean velocities is distinct, e.g., binary granular mixture or fluidized bed.

 We must remind ourselves that the granular binary hydrodynamics equations, as outlined in previous sections of this article, can only deal with the dilute and near elastic granular systems.  We argue that our model\cite{Wang2015193} is closer to the real behavior of dilute granular flow in air-fluidized beds than the previous results.

\section{Acknowledgments}
This work is financially supported by the Ministry of Science and Technology of the People's Republic of China under Grant No. 2012CB215003, by the National Natural Science Foundation of China under Grant Nos. 91334204 and 21176240, by the Chinese Academy of Sciences under Grant No. XDA07080100 and by China Postdoctoral Science Foundation funded project No. 2014M561071.

\section*{References}

\bibliography{apstemplate}

\end{document}